\documentclass[pra,twocolumn,groupedaddress]{revtex4}
\usepackage{graphicx}

\begin{document}

\bibliographystyle{unsrt}

\title{Heralded Two-Photon Entanglement from Probabilistic Quantum Logic Operations on Multiple Parametric Down-Conversion Sources}
\author{T.B. Pittman}
\author{M.M. Donegan}
\author{M.J. Fitch}
\author{B.C. Jacobs}
\author{J.D. Franson}
\affiliation{Johns Hopkins University, Applied Physics Laboratory, Laurel, MD 20723}
\author{P. Kok}
\author{H. Lee}
\author{J.P. Dowling}
\affiliation{Quantum Computing Technologies Group, Section 367, Jet Propulsion Laboratory, California Institute of Technology, Mail Stop 126-347, 4800 Oak Grove Drive, Pasadena, CA 91109}

\date{\today}

\begin{abstract}
An ideal controlled-NOT gate followed by projective measurements can be used to identify specific Bell states of its two input qubits. When the input qubits are each members of independent Bell states, these projective measurements can be used to swap the post-selected entanglement onto the remaining two qubits.  Here we apply this strategy to produce heralded two-photon polarization entanglement using  Bell states that originate from independent parametric down-conversion sources, and a particular probabilistic controlled-NOT gate that is constructed from linear optical elements. The resulting implementation is closely related to an earlier proposal by Sliwa and Banaszek [quant-ph/0207117], and can be intuitively understood in terms of familiar quantum information protocols. The possibility of producing a ``pseudo-demand'' source of two-photon entanglement by storing and releasing these heralded pairs from independent cyclical quantum memory devices is also discussed.
\end{abstract}

\maketitle

The entanglement of two (or more) particles remains one of the most fascinating aspects of quantum mechanics \cite{einstein35}. In addition to its relevance in a number of fundamental issues \cite{bellbook}, quantum entanglement has recently been identified as a valuable resource for a variety of quantum information processing tasks \cite{neilsenchuangbook}. One promising approach to the practical realization of many of these tasks relies on qubits that are encoded in the polarization states of single photons. Consequently, a reliable source of heralded or even ``push-button''\cite{blatt00} two-photon polarization entanglement is of great interest at the present time.

One recent suggestion for such a source involves photon-pair production from controlled biexciton emission of a single quantum dot \cite{benson00,stace02}. Alternatively, one can consider a source based on the photon-pairs produced in parametric down-conversion \cite{klyshkobook}.  However, the random nature of this spontaneous emission process prohibits the direct use of a single two-photon down-conversion source for the production of on-demand pairs, and schemes based on conventional entanglement swapping \cite{zukowski93,pan98} between two down-conversion sources can lead to false heralding signals due to double pair emission from one of the sources \cite{kok00a}. In fact, Kok and Braunstein have shown that the production of one heralded polarization-entangled photon pair using only conventional down-conversion sources, linear optical elements, and projective measurements requires at least three initial pairs \cite{kok00b}.  

In this brief paper, we describe a method for producing heralded two-photon entanglement along theses lines.  We consider the use of our previously proposed probabilistic controlled-NOT gate \cite{pittman01a}, which consumes one entangled photon pair as a resource, to essentially perform an entanglement swapping procedure on two additional entangled photon pairs.  The end result is a unique detection signature of four photons that heralds the presence of one remaining polarization-entangled pair, with a potentially negligible probability of false heralding due to undesired multiple pair emission from the three independent initial pair sources.  As will be shown, the resulting set-up is closely related to an earlier proposal by Sliwa and Banaszek that relies on triple-pair emission from a single down-conversion source \cite{sliwa02}. Other related work includes proposals for generating heralded two-photon polarization entanglement from multiple sources of single photons \cite{hofmann02,zou02,bin02a}, and methods for generating heralded photonic ``path-entanglement'' from single-photon sources \cite{lee02} or down-converted photon pairs \cite{lee00,lombardi02,bin02b}.

\begin{figure}[b]
\vspace*{-.4in}
\includegraphics[angle=-90,width=3in]{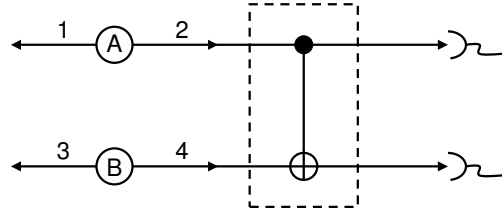}
\vspace*{-.7in}
\caption{An overview of the proposed method to generate heralded two-photon entanglement. $A$ and $B$ are envisioned as independent down-conversion sources, and the controlled-NOT gate, followed by polarization-sensitive single-photon detectors, is used to swap post-selected entanglement of modes $2$ and $4$ onto the remaining photons in modes $1$ and $3$. }
\label{fig:overview}
\end{figure}

Our approach is based on the well-known fact that a controlled-NOT gate, followed by certain measurements, can be used to identify specific Bell states of its two input qubits \cite{cnot}.  As shown in Figure \ref{fig:overview}, we therefore consider the interesting case when the two input qubits are each members of independent Bell states. Specifically, we envision $A$ and $B$ as independent down-conversion sources designed \cite{kwiat95} to produce two-photon Bell states of the form $\phi^{+}_{A}=\frac{1}{\sqrt{2}}[0_{1}0_{2}+1_{1}1_{2}]$, $\phi^{+}_{B}=\frac{1}{\sqrt{2}}[0_{3}0_{4}+1_{3}1_{4}]$.  Assuming that exactly one pair has been emitted by each source, it can be seen that the total four-photon state after the operation of the controlled-NOT gate is of the form \cite{gottesman99}:

\begin{equation}
\psi=\frac{1}{2}[0000+0011+1101+1110]
\label{eq:pre}
\end{equation}

\noindent where the modes are sequentially labelled, 1,2,3,4.  By considering the heralding signal to be the joint detection of a single photon with logical value 0 in mode $4$, and a single photon with logical value $\frac{1}{\sqrt{2}}[0+1]$ (or, say, a Hadamard followed by detection of value 0) in mode $2$, it can be seen from equation \ref{eq:pre} that the heralded output in modes $1$ and $3$ is the Bell state $\phi^{+}_{out}=\frac{1}{\sqrt{2}}[0_{1}0_{3}+1_{1}1_{3}]$.

This measurement procedure can be understood as follows: the detection of a single photon by each (ideal) detector assures that exactly one pair has been emitted from each random down-conversion source, while the requirement that these detected photons have the logical values described above serves to post-select the $\phi^{+}$ Bell state of the controlled-NOT inputs.  Because these inputs are members of $\phi^{+}$ Bell states, this post-selected entanglement is swapped \cite{zukowski93} onto the remaining photons in modes $1$ and $3$.

\begin{figure}[t]
\includegraphics[angle=-90,width=3in]{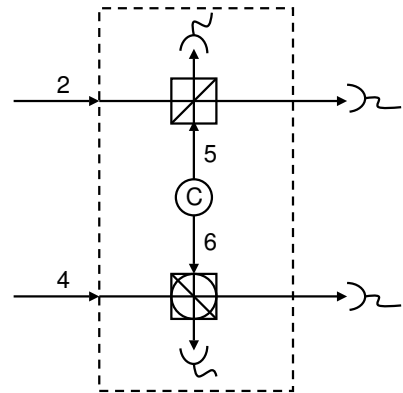}
\caption{The use of our proposed \protect\cite{pittman01a} non-deterministic controlled-NOT gate for the protocol outlined in Figure \protect\ref{fig:overview}. }
\label{fig:aplcnot}
\end{figure}

From a practical point of view, the key feature of this protocol is that a   probabilistic controlled-NOT gate \cite{koashi01,knill01a} (as opposed to a deterministic one) may be sufficient for the production of the heralded entanglement. We therefore consider the implementation of the protocol using  our proposed probabilistic controlled-NOT gate, as shown in Figure \ref{fig:aplcnot}.  The details of this probabilistic quantum logic gate are described in reference \cite{pittman01a}, and some limited aspects of its operation have been experimentally demonstrated \cite{pittman02a}.

To briefly review, our controlled-NOT gate is constructed from two polarizing beam splitters (PBS) which transmit horizontally polarized photons (qubit value 0), and reflect vertically polarized photons (qubit value 1) when aligned in the computational basis.  As seen in Figure \ref{fig:aplcnot}, the upper PBS is aligned in the computational basis, while the lower PBS is rotated by $45^{o}$ and therefore transmits polarization states corresponding to $|+\rangle=\frac{1}{\sqrt{2}}[0+1]$, while reflecting polarization state corresponding to $|-\rangle=\frac{1}{\sqrt{2}}[0-1]$. These two PBS's are linked by a source of two entangled ancilla photons (labelled $C$) which, once again, we envision as a down-conversion source designed \cite{kwiat95} to produce a two-photon Bell state of the form $\phi^{+}_{C}=\frac{1}{\sqrt{2}}[0_{5}0_{6}+1_{5}1_{6}]$.  Given the entangled ancilla state, this controlled-NOT gate signals its successful operation on two input qubits by the joint detection of a single photon in the $+/-$ basis above the upper PBS, and a single photon in the computational basis below the lower PBS \cite{pittman01a}.  The probability of successful operation is $\frac{1}{4}$ provided that feed-forward control \cite{pittman02b} is used to perform various single qubit corrections in the outputs.

Within the framework of the protocol outlined in Figure \ref{fig:overview}, the heralding signal for the desired polarization entanglement in modes $1$ and $3$ would therefore consist of the detection of a single photon by each of the four ideal detectors.  In the first step, the successful operation of the controlled-NOT gate is signalled by the detection of single photons in modes $5$ and $6$. In the second step, the detection of a single photon in each of the controlled-NOT gate outputs is used to perform the post-selection and swapping of the entanglement as described earlier.

One advantage of this particular arrangement is that it is nearly immune to false heralding signals due to undesired multiple pair emission by the three initial pair sources.  For example, double (or even triple) pair emission from either source $A$ or $B$ simply cannot distribute four single photons among the four detectors. Given ideal photon-number resolving detectors, the same is true for double pair emission from any one of the three sources combined with single pair emission from either of the two remaining sources.  

Furthermore, double pair emission from source $C$ alone cannot result in the correct distribution of photons as a consequence of the two PBS's being aligned $45^{\circ}$ with respect to one another.  This latter interference effect can be understood by the fact that amplitudes (expressed in the computational basis) describing photon pairs in mode $5$ must be orthogonally polarized to be correctly distributed by the upper PBS (ie. one into each output port), while those in the mode $6$ must be parallel to have some probability of being correctly distributed by the lower PBS.  In contrast to these requirements, double pair emission from the $\phi^{+}$ Bell state source $C$ is of the form $\frac{1}{2}[0000+0101+1010+1111]$, where the first two qubits correspond to photons in mode $5$ while the second two qubits correspond to photons in mode $6$. In other words, amplitudes that would be successfully distributed by the upper PBS are correlated with amplitudes that can  not be correctly distributed by the lower PBS, and vice-versa.

Even with ideal detectors, however, a false heralding signal could be obtained from simultaneous double pair emission from both source $A$ and source $B$. 
Although the probability of this quadruple pair process can be made significantly smaller than the desired triple pair emission process in the limit of weak down-conversion, such contamination of the heralded Bell state may limit the usefulness of this method for applications in which these errors can not be tolerated.

\begin{figure}[t]
\includegraphics[angle=-90,width=3in]{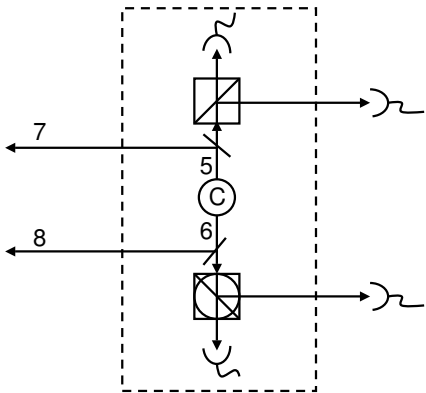}
\caption{A loose interpretation of the Sliwa-Banaszek proposal \protect\cite{sliwa02} in terms of our proposed controlled-NOT gate \cite{pittman01a}.}
\label{fig:sliwa}
\end{figure}

As we mentioned earlier, the overall set-up which results from implementing the general protocol outlined in Figure \ref{fig:overview} with our particular probabilistic controlled-NOT gate \cite{pittman01a} is closely related to an earlier proposal by Sliwa and Banaszek \cite{sliwa02}. For comparative purposes, a loose interpretation of that proposal in terms of our probabilistic controlled-NOT gate is shown in Figure \ref{fig:sliwa}.  Roughly speaking, additional (non-polarizing) beam splitters are  inserted in modes 5 and 6, and the required three initial photon pairs are generated by triple-pair emission from the single source $C$.  In this configuration, the heralding signal (of four single-photon detections) is analogous to that described above, and the production rate of heralded-pairs (in modes $7$ and $8$) was found to be optimized when the reflectivity of the additional beam splitters is $\frac{2}{3}$ \cite{sliwa02}.

Finally, it is instructive to consider the conversion of a heralded source of two-photon polarization entanglement, produced by any means, into a source of entanglement on ``pseudo demand''.   A pseudo-demand source approximates a ``push-button'' source in the sense that the heralding signal is used to activate a storage mechanism for the heralded resource, which can then be released on-command at some chosen later time. Pseudo-demand sources of this kind are particularly relevant in the optical realization of quantum information protocols that can operate on a fixed clock cycle such as that provided by the repetition rate of a master pulsed laser. For these applications the storage mechanisms can be based on the use of simple repetitive optical loops whose round-trip propagation times are equal to the period of the clock cycle.

\begin{figure}[b]
\includegraphics[angle=-90,width=3in]{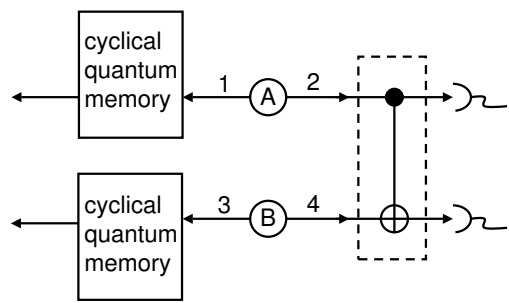}
\vspace*{-.5in}
\caption{Conversion of the heralded source of two-photon entanglement from Figure \ref{fig:overview} into a source of entanglement on ``pseudo demand''.  The heralding signal provided by joint photon detections in the output of the controlled-NOT gate is used to activate cyclical quantum memory devices \protect\cite{pittman02d} in the two modes (1 and 3) containing the heralded entanglement. The envisioned periodic process involves producing and storing the entanglement in advance, and releasing it when needed. }
\label{fig:pseudodemand}
\end{figure}

We previously demonstrated the conversion of a heralded single-photon source into a source of single-photons on pseudo demand in this manner \cite{pittman02c}.  In that experiment, the detection of one member of a down-conversion photon pair heralded the presence of its twin photon, and activated a switching mechanism to reroute the twin into a storage loop. By analogy, the conversion of a source of heralded two-photon polarization entanglement into a pseudo-demand source of entanglement can be accomplished by storing each member of the heralded pair in its own
storage loop, as shown in Figure \ref{fig:pseudodemand}.  The storage loops and associated switching mechanisms, however, must not measure or alter the value of their stored qubits in any way, and in this sense are equivalent to a cyclical (or periodic) quantum memory device. A proof-of-principle demonstration of one particular implementation of a suitable cyclical quantum memory device for single-photon qubits is described in reference \cite{pittman02d}.

Whereas loss in either of the quantum memory devices will cause an obvious reduction in the quality of the stored and released Bell state, the effects of phase shifts are not as detrimental. In fact, 
because one member of the heralded pair is contained in each cyclical quantum memory device of Figure \ref{fig:pseudodemand}, overall phase shifts between the two devices essentially factor out of the two-photon state in such a way that phase-locking of the two memory loops is not required to maintain the coherence of the stored Bell state.  However, any relative phase shifts between qubit values 0 and 1 (ie. birefringence for the case of polarization encoded qubits) in the individual memory devices will result in a final state of the form  $\frac{1}{\sqrt{2}}[0_{1}0_{3}+ e^{i\phi}1_{1}1_{3}]$, where $\phi$ represents the total of these phase shifts.  One of the tangible benefits of the particular cyclical quantum memory device described in reference \cite{pittman02d} is a robustness against these types of relative phase-shift errors due to repeated bit-flipping of the stored qubits.

In summary, we have described a protocol for producing heralded two-photon polarization entangled states by performing probabilistic quantum logic operations on multiple parametric down-conversion sources.  We then described a method for converting a source of heralded entanglement into a source of entanglement on pseudo demand by controlled storage of the entangled photon pairs in two independent cyclical quantum memory devices \cite{pittman02d}.  Such a pseudo-demand source of entanglement is expected to be valuable for the optical realization of a variety of quantum information processing tasks, particularly within the framework of linear optics quantum computation \cite{knill01a,franson02a}.

For example, the primary motivation for this work was to provide a 
pseudo-demand source of two-photon polarization entanglement to use as the entangled ancilla pair required for a complete (as opposed to post-selected) demonstration of our previously proposed non-deterministic controlled-NOT gate \cite{pittman01a}.  It is interesting to see that the controlled-NOT gate itself, run with a source of random entanglement, can be used to produce the deterministic ancilla entanglement needed for a complete implementation of the gate.  Such ``bootstrapping'' of resources and lower efficiency logic gates to produce higher efficiency gates is a fundamental feature of linear optics quantum computation \cite{knill01a}.

    An experimental implementation of the protocol described in this paper presents a formidable challenge in several regards. By far the most significant difficulty appears to be the implementation of the high efficiency photon-number resolving detectors that are required to reject certain false heralding signals during the post-selection procedure.  In addition, the suppression of a false heralding signal due to double-pair emission from source $C$ is based on an experimentally delicate destructive four-photon interference effect. Because an error of that kind would correspond to double-pair emission while the effects of interest correspond to triple-pair emission, the visibility of this destructive interference would have to be very near unity. Although the required mode-matching would be facilitated to some degree by the fact that source $C$ is a single down-conversion crystal \cite{ou99,eibl03}, so far the highest reported visibilities for any type of multi-photon interference effects have only been in the range of roughly $0.85$ to $0.95$ (see, for example, \cite{lamaslinares01,jennewein02,deriedmatten02,pittman03,pan03,zhao03}).

TBP, MMD, MJF, BCJ, and JDF were supported by Army Research Office,  National Security Agency, Advanced Research and Development Activity, Office of Naval Research, and Independent Research \& Development funding. JPD, PK, and HL would like to acknowledge that this work was performed in part by the Jet Propulsion Laboratory, California Institute of Technology, under a contract with the National Aeronautics and Space Administration, with additional support by the Office of Naval Research, the Defense Advanced Research Projects Agency, the National Security Agency, the Advanced Research and Development Activity, and the National Reconnaissance Office.



\end{document}